\documentclass[a4paper]{ISOStyle}
\usepackage{graphicx}

\begin{document}

\title{Classification of ISO SWS\,01 spectra of proto-planetary nebulae: a 
search for precursors of planetary nebulae with [WR] central stars.
}

\author{Ryszard Szczerba\inst{1}, Gra{\.z}yna Stasi{\'n}ska\inst{2},
Natasza Si{\'o}dmiak\inst{1} \and S{\l}awomir K. G{\'o}rny\inst{1}} 
  
\institute{N. Copernicus Astronomical Center, Rabia{\'n}ska 8, 
87-100 Toru{\'n}, Poland
  \and 
LUTH, Observatoire de Paris-Meudon, 5 place Jules Jansses, 92150 Meudon, France
}

\maketitle 

\begin{abstract}
We have analyzed ISO SWS\,01 observations for 61 proto-planetary 
~nebulae ~candidates ~and ~classified ~their spectra according to their 
dominant chemistry. On the basis of our classification and the more general 
classification of SWS\,01 spectra by Kraemer et al.\,(2002) we discuss the 
connection between  proto-planetary nebulae candidates and planetary 
nebulae, with emphasis on possible precursors of planetary nebulae 
with [WR] central stars.

\end{abstract}

\section{INTRODUCTION}

The proto-planetary nebula (PPN) phase, in the context of the late stages of 
low- and intermediate-mass star evolution, is a short lasting period 
(of the order of at most a few thousands of years) when stars 
evolve from Asymptotic Giant Branch (AGB) to planetary nebula (PN). The PPN 
phase is characterized by a rather small mass loss rate 
($\sim$10$^{-7}$\,M$_{\odot}$yr$^{-1}$), 
a decrease of the circumstellar shell optical 
depth due to expansion, and a decrease of the star radius and consequent
increase of the effective temperature (typically from about 4000\,K until the
onset of ionization at about 25000\,K) due to gradual consumption of
stellar envelope, both by nuclear processing and by stellar wind. 

PPNe, being immediate precursors of PNe deserve special 
attention because they offer the possibility to identify the main physical and 
chemical processes which lead to diversity of shapes and chemical 
compositions in PNe and their central stars. One of the most intriguing class 
of PNe is the class of planetary nebulae with Wolf-Rayet type central 
stars (hereafter [WR]PNe -- see e.g. Tylenda et al.\,1993, 
G\'orny \& Stasi\'nska\,1995, Tylenda\,1996, Leuenhagen \& Hamann\,1998). 
Among about 1500 galactic PNe $\sim$50 are [WR]PNe (G\'orny \& Tylenda\,2000).
Observations with the Infrared Space Observatory (ISO, Kessler 
et al. 1996) have shown that in [WR]PNe both forms of dust 
(C-rich: PAHs and O-rich: crystalline 
silicates) are present (Waters et al.\,1998a, Cohen et al.\,1999). For a 
discussion of scenarios put forward to explain this unexpected 
discovery see recent papers by Cohen et al. (2002), De Marco et al.\,(2002) 
and De Marco \& Soker\,(2002). 

In this paper we report on a search of the ISO Data Archive (IDA) for 
spectra of PPNe taken with the 
Short Wavelength Spectrometer (SWS\,01, de Graauw et al. 1996) 
and present a classification of these spectra. We then briefly discuss 
the relation between PPNe and PNe based on the results of this 
classification and the one performed by Kraemer et al.\,(2002).

\section{SAMPLE AND ISO DATA REDUCTION}

Recently, Szczerba et al.\,(2001b) compiled from the literature a list of 
220 PPNe candidates. We have searched the IDA for SWS\,01 
data within 1 arc-min around the IRAS position (or other position if 
the source has no IRAS name) for all PPNe candidates from the Szczerba's list. 
We have found 83 SWS\,01 spectra for 61 objects.

The ISO SWS\,01 data (offline processing - OLP version 10.1) analyzed in this 
work were all processed using ISAP (ISO Spectroscopic Analysis Package) 
version 2.1. The data analysis consisted of extensive bad data removal 
primarily to minimize the effect of cosmic rays. First, all detectors were 
compared to identify possible features. Then, the best detector was chosen 
to compare one by one with others. Finally, the spectra were averaged, using 
median clipping to discard points that lay more than 2.5$\sigma$ from the 
median flux. Whenever memory effects or irregularities were present in the two
scans of SWS\,01 data, we averaged the two scans separately. Then the 
resulting two sub-spectra were used to check the reality of possible features.
The spectra were averaged typically to a resolution of 300, 500, 800 and 1500 
for SWS\,01 data taken with speed 1, 2, 3 and 4, respectively. For the purpose
of this paper the spectra were truncated at 27.5\,$\mu$m since bands 3E and 4 
have usually poor signal-to-noise ratio. In addition, the memory effects can 
influence our ability to recognize crystalline silicates, while cosmic
rays could produce spurious features. Both these effects are difficult to take
into account using the ISAP software only. This then almost 
precludes the recognition 
of the presence of crystalline silicates from the spectra we 
considered (see discussion below).

\begin{figure}[]
\resizebox{\hsize}{!}{\includegraphics[angle=-90]{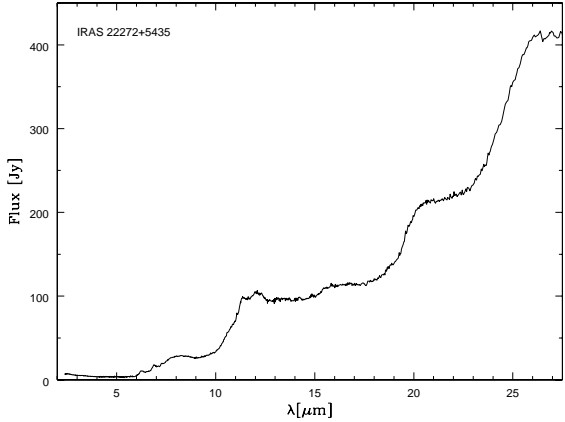}}
\caption{A representative SWS\,01 spectrum (bands 1-3) of PPNe 
with 21\,$\mu$m feature (class C\,21).}
\end{figure}
\begin{figure}[]
\resizebox{\hsize}{!}{\includegraphics[angle=-90]{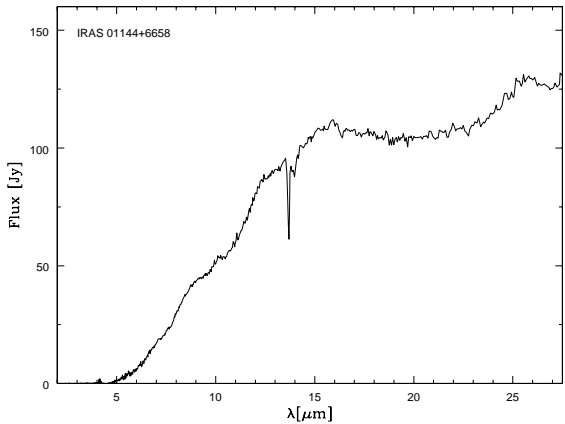}}
\caption{A representative SWS\,01 spectrum (bands 1-3) of PPNe 
with C$_2$H$_2$ and HCN 13.8-14\,$\mu$m features (class C\,mol).}
\end{figure}
\begin{figure}[]
\resizebox{\hsize}{!}{\includegraphics[angle=-90]{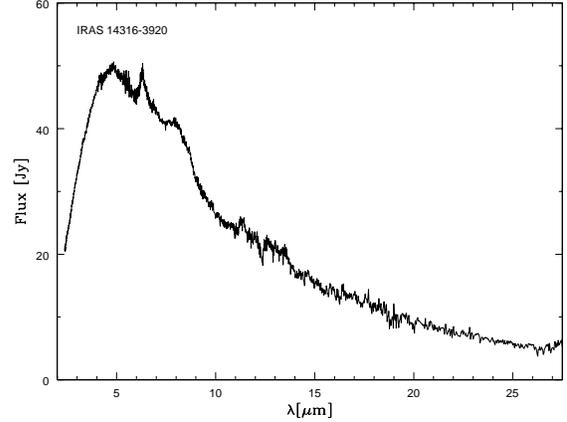}}
\caption{A representative SWS\,01 spectrum (bands 1-3) of R CrB stars 
(class C\,RCrB.)}
\end{figure}
\begin{figure}[]
\resizebox{\hsize}{!}{\includegraphics[angle=-90]{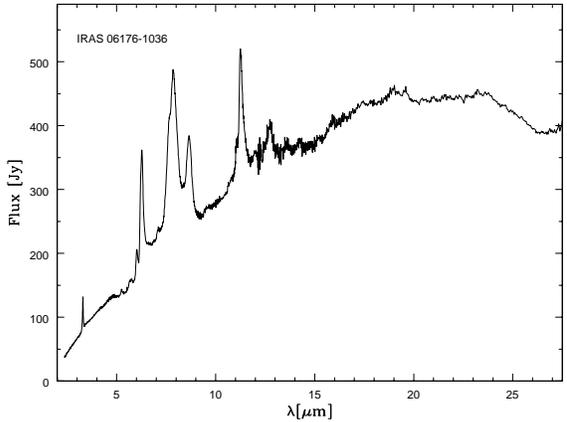}}
\caption{A representative SWS\,01 spectrum (bands 1-3) of PPNe 
with PAH features (class C\,PAH).}
\end{figure}

\section{RESULTS}

The main criterion applied during the classification of the SWS\,01 spectra for
the 61 PPNe candidates was based on the widely accepted CO 
paradigm (the chemistry depends on the C/O ratio in the ejected material). Some
unexpected discoveries of mixed chemistry among late type stars are still
rather more exceptional than typical. The two main 
groups (C- and O-rich) are easily recognized from the analysis of dust and/or
molecular features. Among sources dominated by C-based chemistry we have 
distinguished PPNe with the 21\,$\mu$m feature (C\,21 class), 
sources with C$_2$H$_2$ and HCN 
absorption features around 13.8-14\,$\mu$m (C\,mol class), R\,CrB type sources 
with their characteristic maximum of emission around 6\,$\mu$m 
(C\,RCrB  class). Finally, all the 
sources {\em only} characterized by PAH features were grouped in 
a so-called C\,PAH class. A 
representative SWS\,01 spectrum for each of these groups is shown in 
Figs.\,1-4. Concerning O-rich sources, we simply distinguished between 
PPNe with the 9.7\,$\mu$m 
feature in absorption (Si\,A class) or in emission (Si\,E class), 
and added a class of  RV Tauri type sources where the emission feature
around 10\,$\mu$m has a peculiar shape (Si\,RVTau class). 
Representative SWS\,01 spectra of O-rich PPNe are shown in Figs.\,5-7. 
We were able to classify in such a way 47 sources out of 61 (the spectra of 
the remaining ones are too peculiar to determine what is their dominant 
chemistry).  
Table\,1 shows the classes we assigned for these 47 sources, together 
with the KSPW class (Kraemer et al.\,2002).  Inside each of our 
classes, the sources are 
ordered by decreasing IRAS 25 and 12\,$\mu$m flux ratio
($c21$=F$_{25}$/F$_{12}$). It can be seen that both classifications 
agree rather well. Since Kraemer et al.\,(2002) used {\em automatically}
reduced data from the IDA, their classification is not always accurate. On the 
other hand, as mentioned in Sect.\,2, our classification does not take into 
consideration the possible presence of crystalline silicate features.
\begin{figure}[]
\resizebox{\hsize}{!}{\includegraphics[angle=-90]{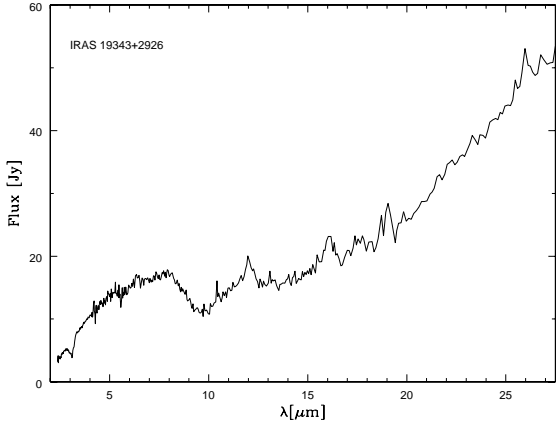}}
\caption{A representative SWS\,01 spectrum (bands 1-3) of PPNe 
with 9.7\,$\mu$m feature in absorption (class Si\,A).}
\end{figure}
\begin{figure}[]
\resizebox{\hsize}{!}{\includegraphics[angle=-90]{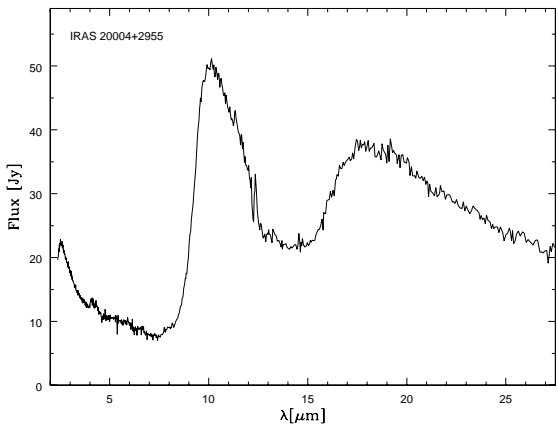}}
\caption{A representative SWS\,01 spectrum (bands 1-3) of PPNe 
with 9.7\,$\mu$m feature in emission (class Si\,E).}
\end{figure}
\begin{figure}[]
\resizebox{\hsize}{!}{\includegraphics[angle=-90]{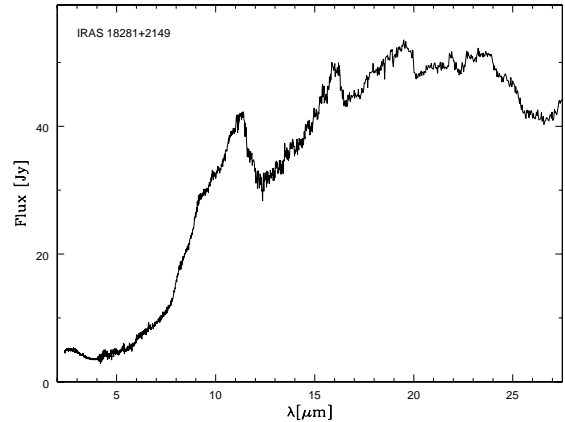}}
\caption{A representative SWS\,01 spectrum (bands 1-3) of RV Tau star
(class Si\,RVTau).}
\end{figure}

\section{DISCUSSION}

%
%
\begin{table}[tb]{}
   \caption[]{Source classification}
  \begin{center}
  \leavevmode
  \footnotesize
    \begin{tabular}{l @{   } l @{  } l | l @{  } l @{   } l }\hline
IRAS         & KSPW$^1$      & {\it c21} & IRAS         & KSPW      & {\it c21} \\
name         & class     &           & name         & class     &  \\
            \hline
            \noalign{\smallskip}
\multicolumn{3}{}{}C\,21 \\
            \noalign{\smallskip}
\object{16594$-$4656}  & 4.CT & 7.72 & \object{20000$+$3239}  & 4.CT  & 3.68 \\
\object{23304$+$6147}  & 4.CT & 7.13 & \object{Z02229$+$6208} & 4.CT  & 2.96 \\
\object{19500$-$1709}  & 4.CT & 5.18 & \object{22272$+$5435}  & 4.CT  & 2.94 \\
\object{07134$+$1005}  & 4.CT & 5.14 & \object{05341$+$0852}  & 4.PN  & 1.79 \\
\object{22574$+$6609}  & 4.PUp:    & 4.09 &               &           &      \\
            \noalign{\smallskip}
\multicolumn{3}{}{}C\,mol \\
            \noalign{\smallskip}
\object{RAFGL 2688}$^2$& 4.CN & 7.15 & \object{01144$+$6658} & 4.CR & 1.79 \\
\object{23321$+$6545} & 4.CN: & 4.69 & \object{19548$+$3035} & 4.CR & 1.57 \\
\object{19480$+$2504} & 4.F   & 3.17 &              &           &      \\
            \noalign{\smallskip}
\multicolumn{3}{}{}C\,RCrB \\
            \noalign{\smallskip}
\object{15465$+$2818} & 3.W  & 0.45 & \object{14316$-$3920} & 3.W   & 0.28 \\
\object{19132$-$3336} & 3.W  & 0.29 &              &           &      \\
            \noalign{\smallskip}
\multicolumn{3}{}{}C\,PAH \\
            \noalign{\smallskip}
\object{13428$-$6232} & 4.Fu   & 15.1 & \object{13416$-$6243} & 4.CN:u & 3.34 \\ 
\object{01005$+$7910} & 4.PUp  & 4.66 & \object{06176$-$1036} & 4.U/SC & 1.24 \\
\object{17347$-$3139} & 4.UE   & 4.62 & \object{16235$-$4832} & 3.CR:: & 0.74 \\
\object{16279$-$4757} & 4.U/SC & 3.82 & \object{10158$-$2844} & 2.U    & 0.21 \\
            \noalign{\smallskip}
\multicolumn{3}{}{}Si\,A \\
            \noalign{\smallskip}
\object{15553$-$5230} & 4.SA: & 7.01 & \object{15452$-$5459} & 4.SAe & 3.55 \\
\object{18276$-$1431} & 4.SA  & 6.45 & \object{17195$-$2710} & 4.SA  & 2.97 \\
\object{17150$-$3224} & 4.SA  & 6.43 & \object{19386$+$0155} & 4.SB  & 2.57 \\
\object{22036$+$5306} & 5.SA  & 5.49 & \object{19343$+$2926} & 5.SA  & 2.44 \\
\object{18596$+$0315} & 5.SA: & 3.84 & \object{17516$-$2525} & 4.SAp & 1.89 \\
            \noalign{\smallskip}
\multicolumn{3}{}{}Si\,E \\
            \noalign{\smallskip}
\object{12175$-$5338} & 4.SE:  & 7.98 & \object{11385$-$5517} & 5.SE: & 1.49 \\
\object{18095$+$2704} & 4.SEC  & 2.79 & \object{20004$+$2955} & 2.SEc & 3.55 \\
\object{18062$+$2410} & 4.SE:  & 2.59 & \object{17534$+$2603} & 3.SEp & 0.46 \\
\object{19244$+$1115} & 4.SEC  & 1.71 &              &            &      \\
            \noalign{\smallskip}
\multicolumn{3}{}{}Si\,RVTau \\
            \noalign{\smallskip}
\object{18281$+$2149} & 4.SEC  & 1.49 & \object{18448$-$0545} & 2.SEa: & 0.45 \\
\object{22327$-$1731} & 4.SE:: & 1.30 & \object{12185$-$4856} & 7  R   & 0.35 \\ 
\object{20117$+$1634} & 4.SE:: & 0.56 &              &            &     \\
            \noalign{\smallskip}
            \hline
\multicolumn{2}{}{}
$^1$--Kraemer et al.\,(2002);\\
\multicolumn{5}{}{}
$^2$--RAFGL\,2688 has not been observed by IRAS.\\
         \end{tabular}
  \end{center}
\end{table}

From Table\,1, the proportion of C-rich  sources is 25/47 (53\,\%). However, 
because of source selection, the ISO sample is biased towards sources with 
the 21\,$\mu$m feature, which amount to 9 objects in our sample. Therefore, 
the unbiased proportion of C-rich PPNe candidates is somewhere between 
16/38 (42\,\%) and  53\,\%. Interestingly, the proportion of C-rich PNe 
is 35--40\,\% (Rola \& Stasi\'nska\,1994, Kingsburgh \& Barlow\,1994). 
This suggests that most PPNe candidates from our sample will indeed become 
PNe (from now on we will drop the word ``candidate'' for simplicity).

We can now roughly estimate what number of PPNe from our classified 
ISO sample are expected to become [WR]PNe.  G{\'o}rny \& Stasi{\'n}ska\,(1995) 
argued that the proportion of [WR]PNe relative to the total number of PNe
is about 8\,\%. Then one expects the same proportion of [WR]PNe 
precursors among the sample of PPNe. This implies that 8\,\% of 47 
PPNe, i.e $\sim$4 will become [WR]PNe. 

As discussed recently by De Marco \& Soker\,(2002) the most important 
characteristic of [WR]PNe seems to be dual dust chemistry.  About 
80\,\% of {\em late} [WR]PNe show the presence of silicates -- 
mostly crystalline -- 
while at the same time all of them show PAH emission, see Szczerba et 
al.\,(2001a). Very likely the dual dust chemistry is a result of O-rich dust 
being formed when the star was O-rich and stored in some stable reservoir 
in the stellar vicinity, while the carbon chemistry 
occurs later when the [WR]PN progenitor is already C-rich. As concerns 
{\em early} [WR]PNe, crystalline silicates are seen in only one object 
(NGC\,5315 -- K. Volk, private communication) out of 6 with available ISO 
spectra. The reason for such a small proportion is not clear. G{\'o}rny \& 
Tylenda\,(2000) argued that there is an evolutionary link between late and 
early [WR]PNe, so a stable  reservoir of crystalline silicates should be seen 
in later phases as well, unless a fast wind is able to destroy or remove the 
crystalline material from the star surroundings. Non-detection of crystalline 
silicates can also be  due to a worse quality of ISO SWS data for early 
[WR]PNe. From the above considerations, it is natural to think that PPNe 
showing dual dust chemistry should evolve into [WR]PNe. In our sample, the 
famous Red Rectangle (IRAS\,06176$-$1006) and  IRAS\,16279$-$4757  contain 
both C-rich dust and crystalline silicates (from the KSPW classification, see 
also Waters et al.\,1998b and Molster et al.\,1999) which makes them good 
candidates to be precursors of late [WR]PNe. Note that not all [WR]PNe are 
C-rich (G\'orny \& Stasi\'nska 1995, De Marco\,2002), in consequence not all 
PPNe that will become [WR]PNe are necessarily C-rich. Therefore, further 
candidates for [WR] PNe precursors in our sample are then the O-rich sources 
AC\,Her (IRAS\,18281$+$2149), IRAS\,18095$+$2704 and IRAS 19244$+$1115.

If non-detection of dual dust chemistry in some [WR] PNe is not due to 
observational effects but is a sign of real absence of crystalline silicates,
then it is possible that the precursors of these [WR]PNe may also not
show crystalline silicate features. Recently, Hony et al.\,(2001) 
reported that some [WR]PNe show the 21\,$\mu$m feature suggesting
a possible link between them and PPNe with the 21\,$\mu$m feature (our 
class C\,21 or KSPW class CT). To our knowledge none of the C\,21 source 
shows evidence of dual dust chemistry so they could be candidates for 
precursors of [WR]PNe  without crystalline silicates. Demographic arguments 
show that this could be the case for only a small fraction of them. Indeed, 
there are at least 12 PPNe with the 21\,$\mu$m feature (see e.g. Kwok et 
al.\,2002) among 220 known PPNe, i.e. 5.5\%. On the other hand only 20\,\% of 
late type [WR]PNe with analyzed ISO spectra do not have crystalline 
silicates. According to G\'{o}rny \& Tylenda\,(2000), 30\,\% of all [WR]PNe 
are of late type and, as mentioned above, [WR]PNe represent 8\% of the total 
number of PNe. Thus, the proportion of {\em late} [WR]PNe without crystalline 
silicates is {\em at most} 20\%$ \times$ 30\%$ \times$ 8\% = 0.5\% of the 
total population of PNe. Therefore, most  of 21\,$\mu$m sources will not go 
through the  {\em late} [WR]PN phase. One could still argue that they could 
evolve into PNe in which the [WR] phenomenon will appear only later. 
However, G\'{o}rny \& Tylenda (2000) have shown that most [WR]PNe do evolve  
from late to early type. Besides, among PNe, the 21\,$\mu$m feature has been 
seen only in [WR]PNe, meaning that PPNe with the 21\,$\mu$m feature cannot 
evolve to early [WR]PNe that have not gone through a late [WR] stage. 
Therefore, PPNe with the  21\,$\mu$m feature cannot be considered, 
as a class, as precursors of [WR]PNe. 
 
\begin{acknowledgements}

This work has been partly supported by grant 2.P03D.024.18p01 of the Polish 
State Committee for Scientific Research, by the Polonium program 
(contract No. 03242XJ) and by the Jumelage France-Pologne.

\end{acknowledgements}

\end{document}